\documentclass[runningheads]{CMSIM}
\usepackage{graphicx} 
\usepackage{amsmath,amssymb}                      
\usepackage[square,sort,comma,numbers]{natbib}

\renewcommand{\thefootnote}

\title*{Reduced-order modeling of the fluidic pinball}

\titlerunning{\it Fluidic pinball}

\author{
Luc R. Pastur\inst{1}
\and
  Nan Deng\inst{2}
  \and 
  Marek Morzy\'nski\inst{3}
  \and 
 Bernd R. Noack\inst{2,4,5}
}

\authorrunning{\it Pastur et al.}

\institute{
IMSIA --- ENSTA ParisTech, 828 Bd des Mar\'echaux, F-91120 Palaiseau, France\\
(E-mail: {\tt luc.pastur@ensta-paristech.fr})
\and
  LIMSI-CNRS, Universit\'e Paris Sud, Universit\'e Paris-Saclay, F-91405 Orsay, France\\
  (E-mail: {\tt nan.deng@u-psud.fr})
   \and 
 Institute of Combustion Engines and Basics of Machine Design, Pozna\'n University of Technology, PL 60-965 Pozna\'n, Poland\\
  (E-mail: {\tt marek.morzynski@put.poznan.pl})
  \and 
  Institute for Turbulence-Noise-Vibration Interaction and Control, Harbin Institute of Technology, Shenzhen, People's Republic of China \\
  \and 
Institut f\"ur Str\"omungsmechanik und Technische Akustik (ISTA), Technische Universit\"at Berlin, D-10623 Berlin, Germany\\
  (E-mail: {\tt noack@limsi.fr})
}

\begin{document}
\thispagestyle{empty}
\maketitle             
\setlength{\leftskip}{0pt}
\setlength{\headsep}{16pt}
\footnote{\begin{tabular}{p{11.2cm}r}
\small {\it $11^{th}$CHAOS Conference Proceedings, 5 - 8 June 2018, Rome, Italy} \\  
   \small \textcopyright {} 2018 ISAST & \includegraphics[scale=0.38]{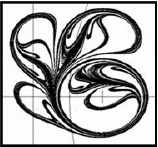}
 \end{tabular}}
\begin{abstract}
The fluidic pinball is a geometrically simple flow configuration with three rotating cylinders on the vertex of an equilateral triangle. Yet, it remains physically rich enough to host a range of interacting frequencies and to allow testing of control laws within minutes on a laptop. The system has multiple inputs (the three cylinders can independently rotate around their axis) and multiple outputs (downstream velocity sensors).
Investigating the natural flow dynamics, we found that the first unsteady transition undergone by the wake flow, when increasing the Reynolds number, is a Hopf bifurcation leading to the usual time-periodic vortex shedding phenomenon, typical of cylinder wake flows, in which the mean flow field preserves axial symmetry. We extract dynamically consistent modes from the flow data in order to built a reduced-order model (ROM) of this flow regime. We show that the main dynamical features of the primary Hopf bifurcation can be described by a non-trivial lowest-order model made of three degrees of freedom. 
\keyword{fluid mechanics, flow control, reduced-order modeling, transition to chaos}
\end{abstract}

\section{Introduction}

Machine learning control (MLC) has been recently successfully applied to closed-loop turbulence control experiments for mixing enhancement \cite{parezanovic2016}, reduction of circulation zones \cite{gautier2015}, separation mitigation of turbulent boundary layers \cite{hu2008a,hu2008b}, force control of a car model \cite{li2016} and strongly nonlinear dynamical systems featuring aspects of turbulence control \cite{brunton2015,duriez2017}. In all cases, a simple genetic programming algorithm has learned the optimal control for the given cost function and out-performed existing open- and closed-loop approaches after few hundreds to few thousands test runs. Yet, there are numerous opportunities to reduce the learning time by avoiding the testing of similar control laws and to improve the performance measure by generalizing the considered control laws. In addition, running thousands of tests before converging to the optimal control law can be out-of-reach when dealing with heavy numerical simulations of the Navier-Stokes equations.  

In order to further improve MLC strategies, it is therefore of the utmost importance to handle numerical simulations of the Navier-Stokes equations in flow configurations that are geometrically simple enough to allow testing of control laws within minutes on a Laptop, while being physically complex enough to host a range of complex dynamical flow regimes. With that aim in mind, Noack \& Morzynski \cite{tutorial} proposed as an attractive flow configuration the uniform flow around 3 cylinders which can be rotated around their axis (3 control inputs), with multiple downstream velocity sensors as multiple outputs. As a standard objective, the control goal could be to stabilize the wake or reduce the drag.
This configuration, proposed as a new benchmark for multiple inputs-multiple outputs (MIMO) nonlinear flow control, was named as the \textit{fluidic pinball} as the rotation speeds allow to change the paths of the incoming fluid particles like flippers manipulate the ball of a real pinball.

With non-rotating cylinders, the steady base flow looses stability, beyond a critical value of the Reynolds number, with respect to an oscillatory vortex-shedding instability. In this flow regime, we show that the reduced order model (ROM) of lowest dimension, though still able to reproduce the dynamical features of the flow regime, has three degrees of freedom. Designing a relevant ROM to describe a complex system is a first step toward the design of winning control strategies, as ROMs both allow testing hundreds to thousands of controllers within a minute and are predictive over a finite time horizon.

\section{The fluidic pinball}

\begin{figure}
  \centerline{
 \includegraphics[width=.75\linewidth]{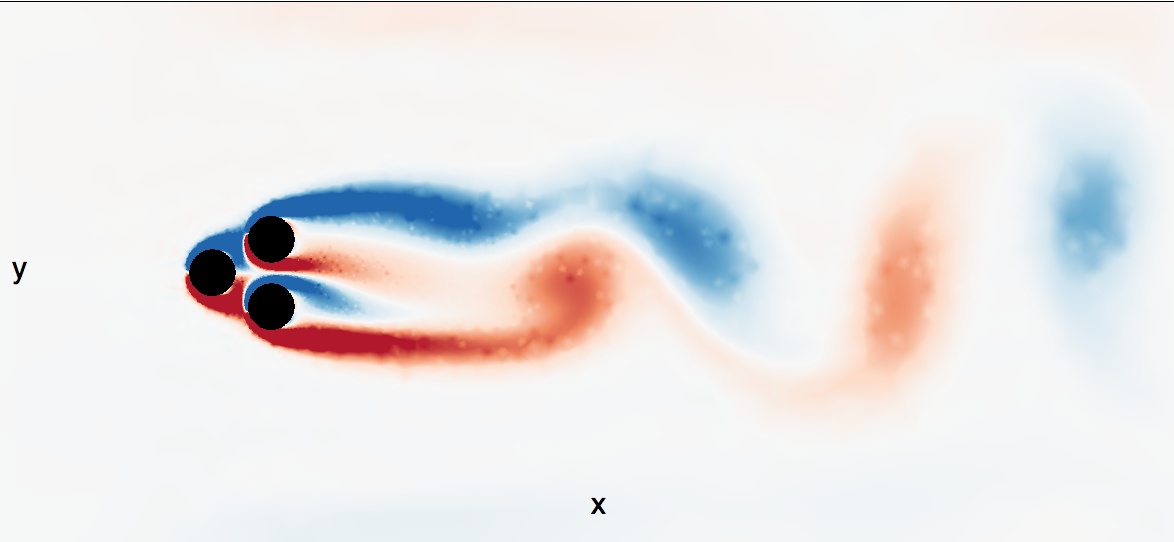}}
 \caption{Configuration of the fluidic pinball: the three cylinders are in black, the flow is coming from the left. The colormap encodes the vorticity field (arbitrary units).  }
 \label{fig:config}
\end{figure}

The \textit{fluidic pinball} is made of three equal circular cylinders of radius $R$ that are placed in parallel in a viscous incompressible uniform flow with speed $U_\infty $. The centers of the cylinders form an equilateral triangle with side-length $3R$, symmetrically positioned with respect to the flow (see Fig.~\ref{fig:config}). The leftmost triangle vertex points upstream, while rightmost side is orthogonal to the oncoming flow. Thus, the transverse extend of the three cylinder configuration is given by $L = 5R$. This flow is described in a Cartesian coordinate system where the $x$-axis points in the direction of the  flow, the $z$-axis is aligned with the cylinder axes, and the $y$-axis is orthogonal to both. The origin 0 of this coordinate system coincides with the mid-point of the rightmost bottom and top cylinder. The location is described by $\mathbf{x} = (x; y; z) = x\,\mathbf{e}_x + y\,\mathbf{e}_y + z\,\mathbf{e}_z$, where $\mathbf{e}_{x;y;z}$ are unit vectors pointing in the direction of the corresponding axes. Analogously, the velocity reads $\mathbf{u} = (u; v;w) = u\,\mathbf{e}_x +v\,\mathbf{e}_y +w\,\mathbf{e}_z$. The pressure is denoted by $p$ and time by $t$. In the following, we assume a two-dimensional flow, i.e. no dependency of any flow quantity on $z$ and vanishing spanwise velocity $w\equiv 0$. The Newtonian fluid is characterized by a constant density $\rho$ and kinematic viscosity $\nu$. In the following, all quantities are assumed to be non-dimensionalized with cylinder diameter $D = 2R$, velocity $U_\infty $ and fluid density $\rho$. The corresponding Reynolds number is defined as $Re_D = U_\infty D/\nu$. The Reynolds number based on the transverse length $L = 5D$ is 2.5 times larger. The computational domain extends from $x = -6$ up to $x = 20$
in the streamwise direction, and from $y = -6$ up to $y = 6$ in the crosswise direction. In these units, the cylinder axes are located at
$$
\begin{array}{lcl}
x_F =  - 3/2 \cos 30^{\circ}, & & y_F = 0, \\
x_B = 0, & & y_B = - 3/4, \\
x_T = 0, & & y_T = + 3/4. \\
\end{array}
$$
Here, and in the following, the subscripts `F', `B' and `T' refer to the front, bottom and top
cylinder. 

The dynamics of the flow is governed by the incompressible Navier-Stokes equations:
\begin{eqnarray}
 \frac{\partial  \mathbf{u}}{\partial t} + \mathbf{u}\cdot \nabla \mathbf{u} & = & -\nabla p + \frac{1}{Re_D} \Delta \mathbf{u},  \\
 \nabla \cdot \mathbf{u} & = & 0, 
 \label{eq:ns}
\end{eqnarray}
where $\nabla $ represents the Nabla operator, $\partial _t$ and $\Delta$ denote the partial derivative and the Laplace operator. 
Without forcing, the boundary conditions comprise a no slip-condition ($\mathbf{u} = 0$) on the cylinder and a free-stream condition ($\mathbf{u} =\mathbf{e}_x$) in the far field. 
The flow can be forced by rotating the cylinders. In the forthcoming part of the paper, however, the cylinders are kept fixed. 

For more details about the numerical setup and the Navier-Stokes solver, the interested reader can refer to the technical report and user manual by Noack \& Morzynski \cite{tutorial}.

\section{Reduced-order model of the vortex shedding flow regime}

\begin{figure}
\centering{
 \includegraphics[width=\linewidth]{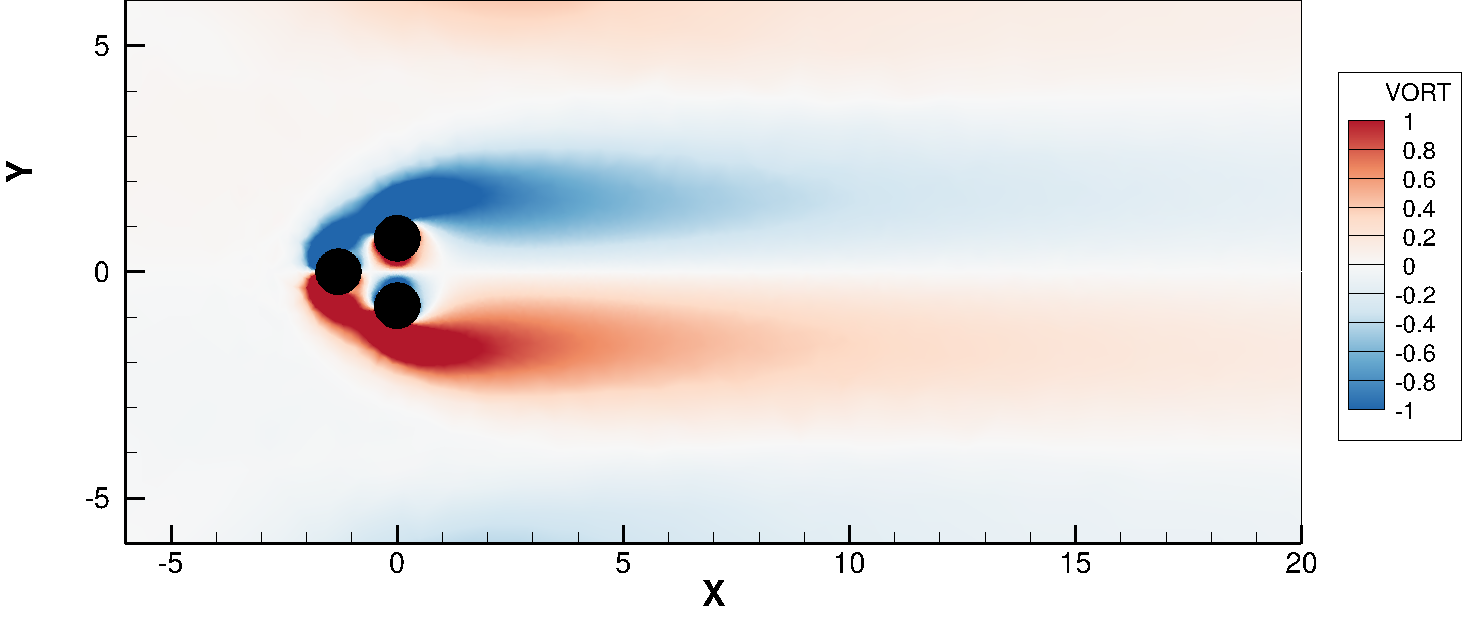}
}
\caption{Steady base flow at $Re_D=10$. The colormap encodes the vorticity field.}
\label{fig:steadysolution}
\end{figure}

\begin{figure}
\centering{
\includegraphics[width=\linewidth]{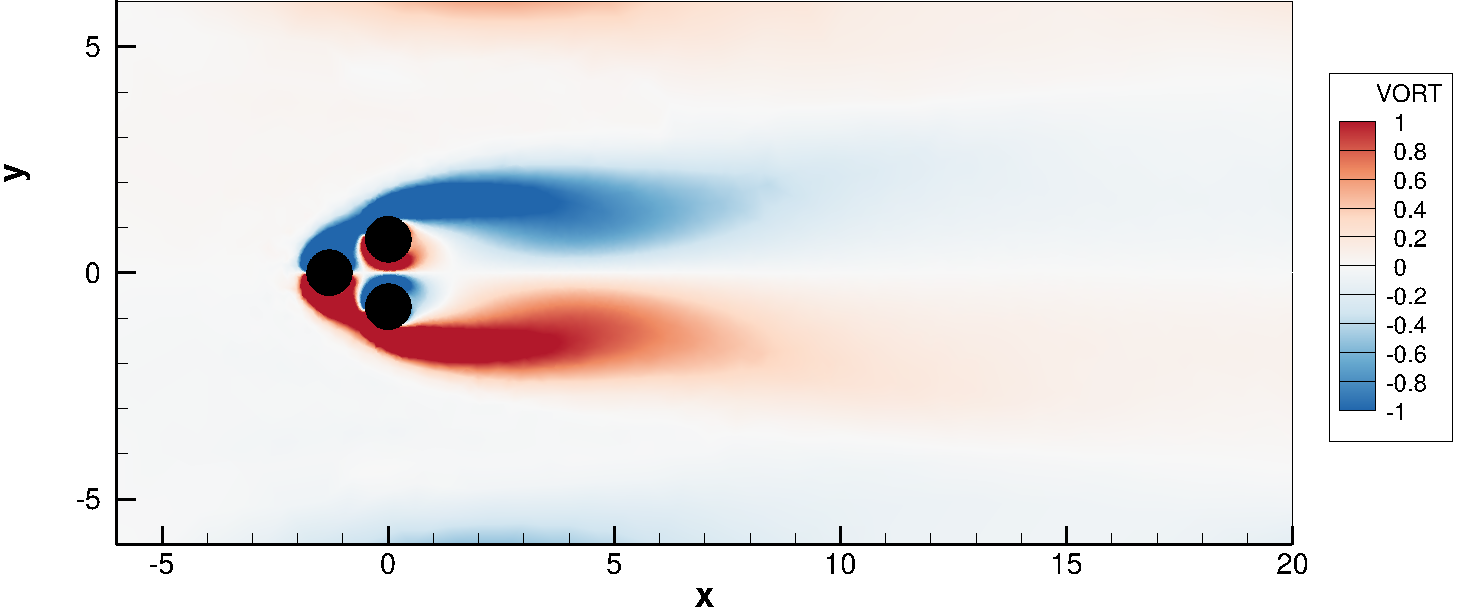}
}
\caption{Mean flow field at $Re_D=30$. The colormap encodes the vorticity field. }
\label{fig:hopf}
\end{figure}

The steady solution, shown in figure~\ref{fig:steadysolution} for $Re_D=10$, is stable up to the critical value $Re_{c}\simeq 18$ of the Reynolds number (the critical value would be about 45 in units of $L$). Beyond this value, the system undergoes a supercritical Hopf bifurcation characterized in the flow field by the usual vortex shedding phenomenon and generation of the von K\'arm\'an vortex street. The associated mean flow field is shown in Fig.~\ref{fig:hopf} for $Re_D=30$. 

\begin{figure}
\centering{
\begin{tabular}{cc}
(a) &
\includegraphics[width=\linewidth]{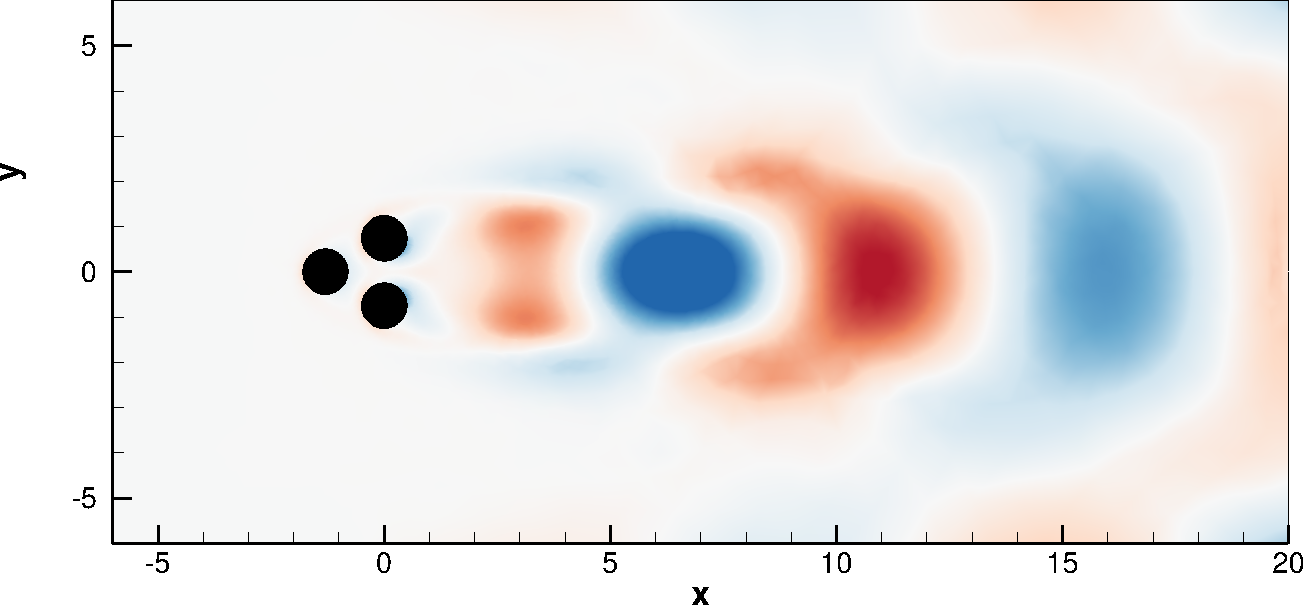} \\
 (b) &
 \includegraphics[width=\linewidth]{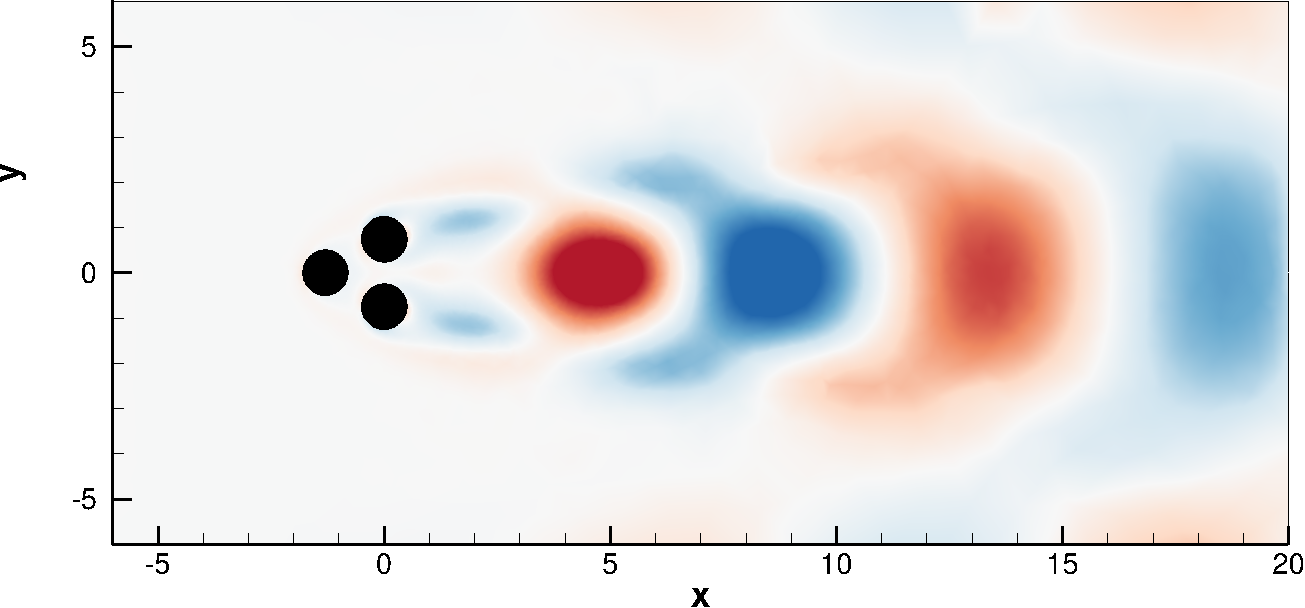} \\
 (c) & 
 \includegraphics[width=\linewidth]{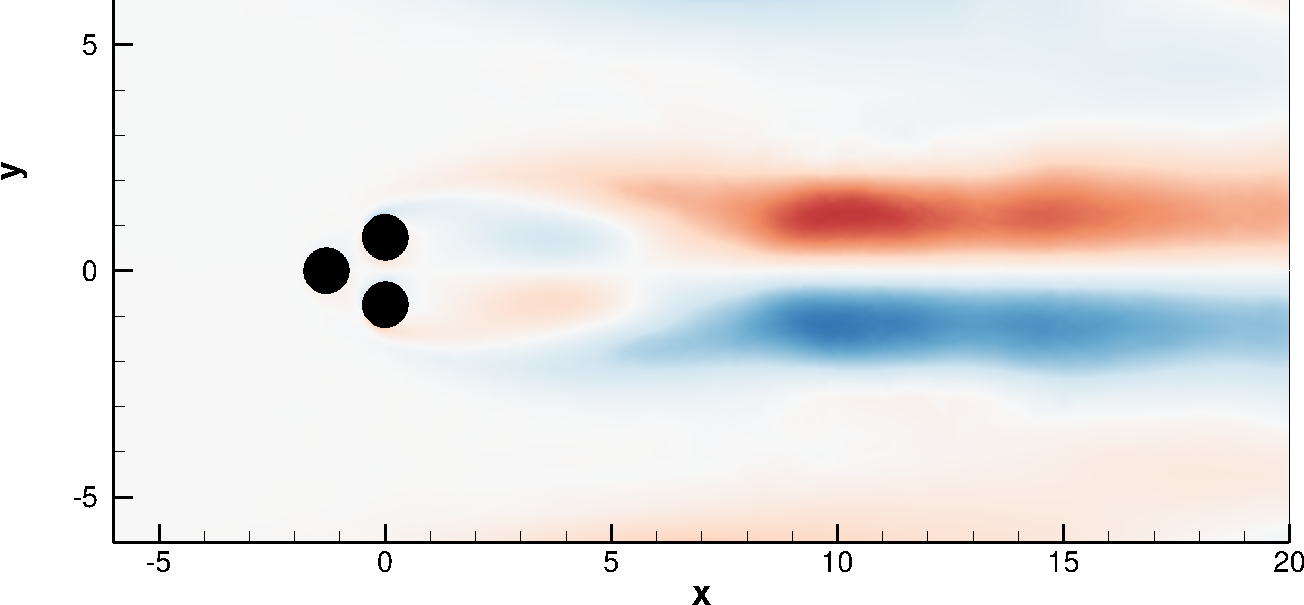} \\
\end{tabular}
}
\caption{First two leading POD modes (a) $\mathbf{u}_1(x,y)$, (b) $\mathbf{u}_2(x,y)$ and (c) shift mode $\mathbf{u}_\Delta (x,y)$, at $Re_D=30$. The colormap encodes the vorticity field. }
\label{fig:u12}
\end{figure}

\begin{figure}
\centering{
\begin{tabular}{cc}
 (a) & (b) \\
 \includegraphics[width=.45\linewidth]{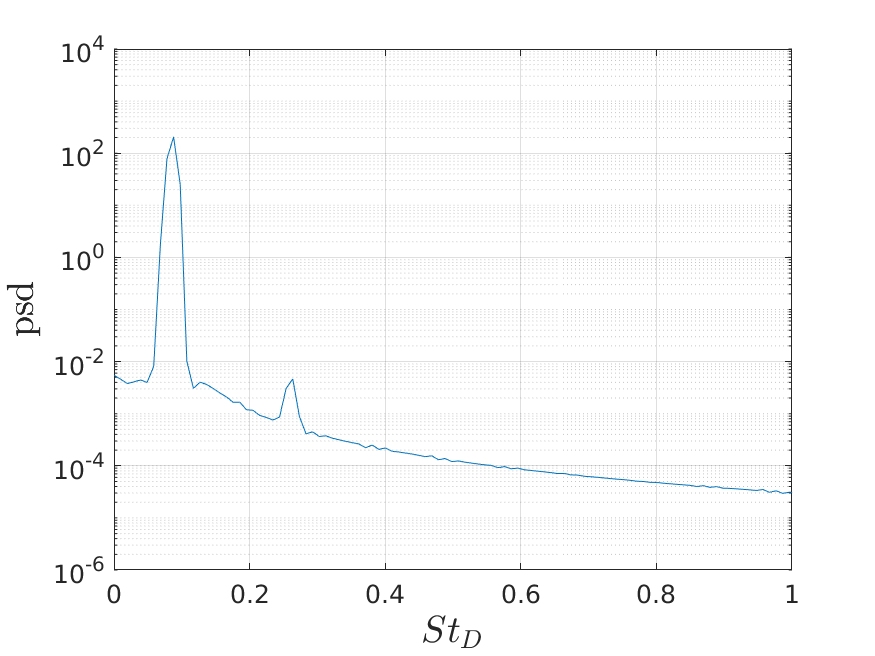} &
 \includegraphics[width=.45\linewidth]{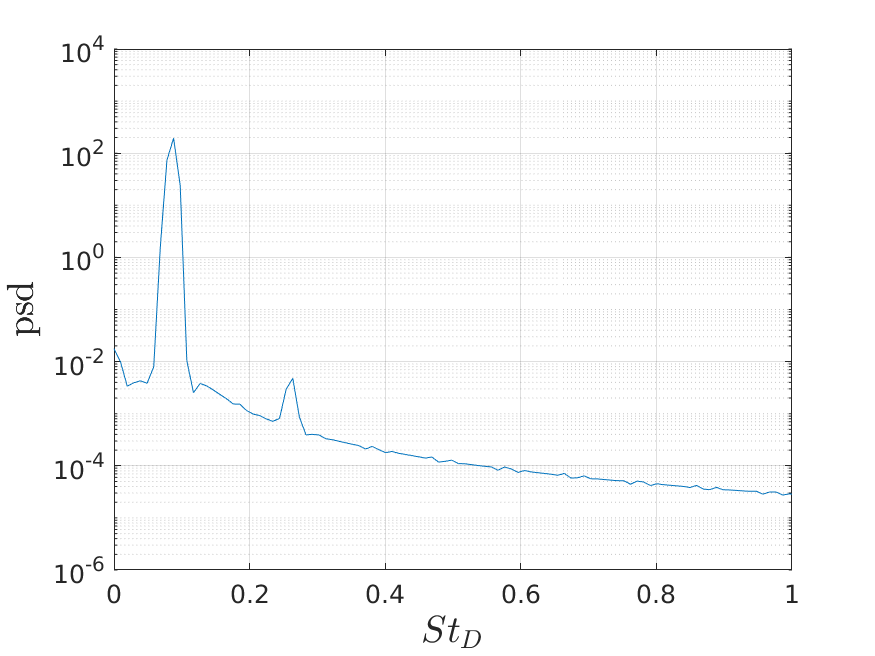} \\
\end{tabular}
}
\caption{Power spectral densities of their associated time coefficients (a) $a_1(t)$ and (b) $a_2(t)$. }
\label{fig:a12}
\end{figure}


Low-dimensional and yet relevant ROMs must rely on the identification of the manifold on which the dynamics takes place. As an illustration, we consider the oscillatory flow regime observed at $Re_D=30$. The inertial manifold hosts both the final oscillatory state and the transient dynamics to the final state. Following \cite{noack2003}, we apply a proper orthogonal decomposition (POD) to the data set made of the fluctuating velocity field, $\mathbf{u}^\prime (x,y;t) = \mathbf{u}(x,y;t) - \bar{\mathbf{u}}(x,y)$, where $\mathbf{u}(x,y;t)$ is the velocity flow field and $\bar{\mathbf{u}}(x,y) =\lim _{T\rightarrow \infty}1/T\int _0^T \mathbf{u}(x,y;t) \, \mathrm{d}t$ is the time-averaged mean flow field. The POD modes $\mathbf{u}_k(x,y)$, $k=1,\hdots, N-1$ with $N$ the number of snapshots $\mathbf{u}(x,y;t)$ in the data set, provide a complete basis of orthogonal modes for the decomposition of any flow field in the data set \cite{berkooz1993}: 
\begin{equation}
 \mathbf{u}(x,y;t) = \bar{\mathbf{u}}(x,y) + \underbrace{\sum _{k=1}^{N-1} a_k(t) \mathbf{u}_k(x,y)}_{\mathbf{u}^\prime(x,y;t)},
 \label{eq:pod}
\end{equation}
where the $a_k$'s are the time coefficients of the decomposition. 
The two leading POD modes $\mathbf{u}_{1,2}(x,y)$ are associated with the vortex shedding phenomenon, as shown in Fig.~\ref{fig:u12}(a)\&(b), together with the power spectral densities of their associated time coefficients $a_1(t)$ and $a_2(t)$ in Fig.~\ref{fig:a12}(a)\&(b), where a dominant peak is found at $St_D = fD/U \simeq 9\times 10^{-2}$ ($St_{L} = fL/U \simeq 0.22$).
Modes $\mathbf{u}_{1,2}(x,y)$, however, are associated with the final oscillatory state around the mean flow field $ \bar{\mathbf{u}}(x,y) $. In order to describe the transient dynamics from the (unstable) steady solution $\mathbf{u}_s(x,y) $ to the final state, it is necessary to introduce as an additional degree of freedom the so-called shift mode $\mathbf{u}_\Delta (x,y)$ defined as $\mathbf{u}_\Delta (x,y) =  \bar{\mathbf{u}}(x,y) -\mathbf{u}_s(x,y)$ and orthonormalized with respect to the leading POD modes \cite{noack2003}. The steady solution $\mathbf{u}_s(x,y)$ is obtained by a Netwon method and the shift mode $\mathbf{u}_\Delta (x,y)$ is shown in Fig.~\ref{fig:u12}(c). Following \cite{noack2003}, let us consider the following truncated flow field:
\begin{equation}
	\tilde{\mathbf{u}}(x,y;t) = \mathbf{u}_s(x,y) + a_\Delta(t)\mathbf{u}_\Delta(x,y) + a_1(t)\mathbf{u}_1(x,y) + a_2(t)\mathbf{u}_2(x,y),
 \label{eq:utilde}
\end{equation}
%
The dynamics of $a_1$, $a_2$, $a_\Delta $ should write:
\begin{eqnarray}
	\dot{a}_1 & = & (\sigma - \kappa_r a_\Delta) a_1 - (\omega +\kappa_i a_\Delta) a_2, \nonumber \\
	\dot{a}_2 & = & (\sigma - \kappa_r a_\Delta) a_2 + (\omega +\kappa_i a_\Delta) a_1, \label{eq:system} \\
	\dot{a}_\Delta & = & -\lambda \left(a_\Delta -  \kappa_\Delta(a_1^2 + a_2^2) \right), \nonumber 
\end{eqnarray}
in order to account for the Hopf bifurcation normal form with triadic interactions between the individual modes, as imposed by the quadratic nonlinearities of the underlying Navier-Stokes equations. 
Identifying the coefficients of the dynamical system \eqref{eq:system} from the transient and final flow regimes, one gets $\sigma = 4.2\times 10^{-2}$, $\omega = 0.5$, $\kappa_r = 1.5\times 10^{-2}$, $\kappa_i = 2.2\times 10^{-2}$,  $\kappa_\Delta = 0.2$ and $\lambda \gg 1$, slaving $a_3$ to $(a_1^2 + a_2^2)$.

\begin{figure}[tb]
\centering{
\includegraphics[width=\linewidth]{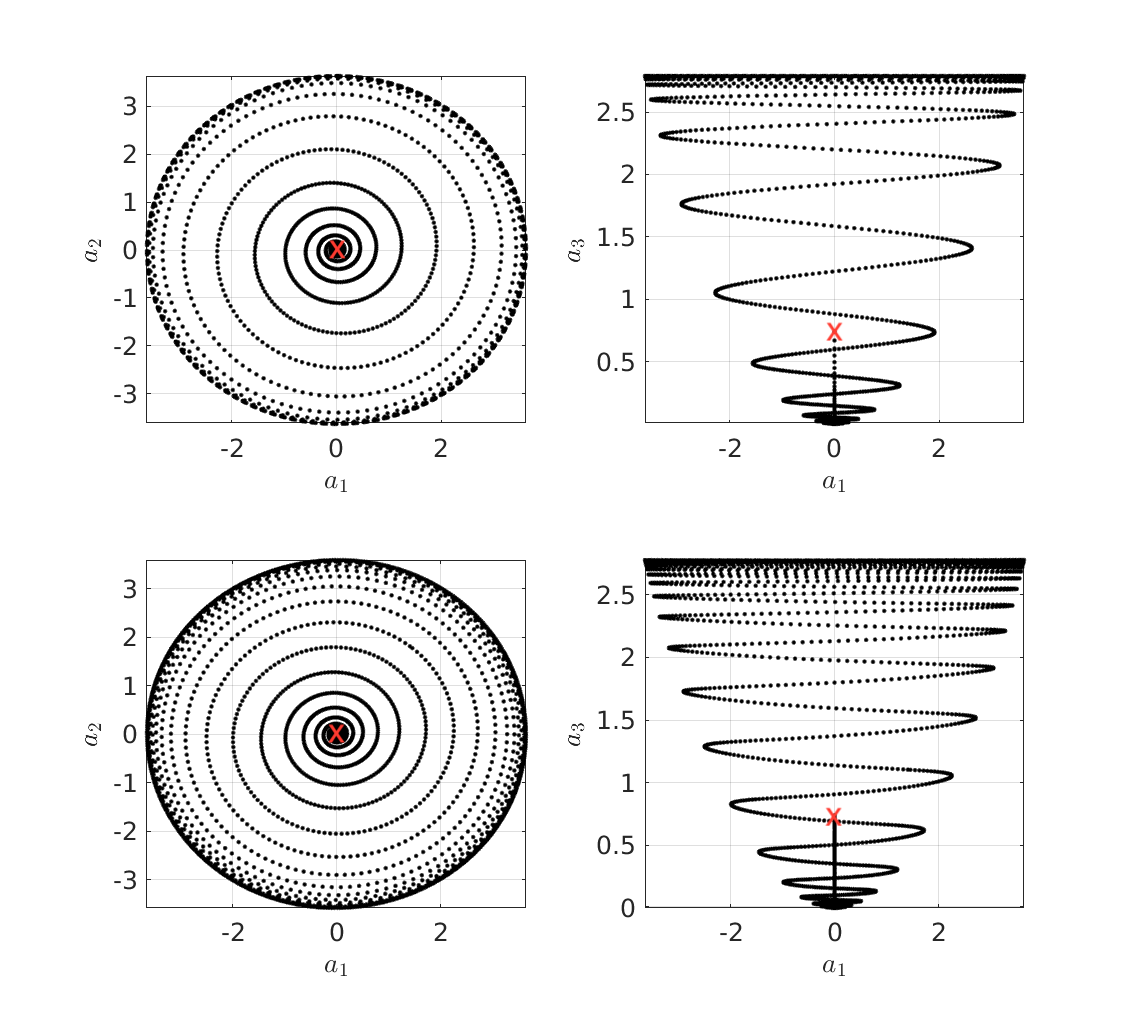}
}
\caption{Phase portraits of the ROM (top) and the fluidic pinball (bottom) from the initial condition (red cross in the figures) to the final oscillatory state (larger limit cycle).}
\label{fig:portrait}
\end{figure}

The dynamics of the ROM from some arbitrary initial condition to the final oscillatory state, integrated with a Runge-Kutta 4.5 numerical scheme, is compared to the dynamics of the fluidic pinball from the same initial condition, see Fig.~\ref{fig:portrait}. In both cases, the final oscillatory states in the phase portraits spanned by $(a_1,a_2)$ are two limit cycles of identical amplitude. In the phase portraits spanned by $(a_1,a_3)$, the parabolic shape of the manifold is identical in the two cases. This means that the inertial manifold on which the dynamics takes place is correctly identified at leading order by our ROM. This also means that the lowest-order model able to reproduce the dynamics of the fluidic pinball, at $Re_D=30$, has at least three degrees of freedom, namely $a_1$, $a_2$, $a_\Delta $, the latter being slaved to the two former. Only the time scales of the transient are not perfectly reproduced, but this should be improved by introducing for instance few additional degrees of freedom or better calibrating $\lambda$. 

\section{Conclusion}

We have considered the fluidic pinball, a newly introduced benchmark configuration for MIMO nonlinear fluid flow control, beyond its primary instability towards the vortex-shedding flow regime. We could propose a reduced-order model based on POD of at least three degrees of freedom that is able to catch the main features of the manifold on which the dynamics takes place. The degrees of freedom are the two leading POD modes, associated with vortex shedding in the final oscillatory state, and, slaved to them, the shift mode that account for the steady solution deformation towards the mean flow field in the final state. 

The ROM was derived for non-rotating cylinders and for a given Reynolds number.
Yet, the fluidic pinball can display a much richer spectrum of dynamic behaviors using three cylinder rotations as free constant parameters.
Thus, one flow configuration allows to reproduce many known nonlinear behaviors.


\section*{Acknowledgment}
This work is part of a larger project involving S. Brunton, G.~Cornejo Maceda, J.C. Loiseau, F. Lusseyran, R.~Martinuzzi, C.~Raibaudo, R.~Ishar and many others. 

This work is supported by the ANR-ASTRID project ``FlowCon'', by a public grant overseen by the French National Research Agency (ANR) as part of the ``Investissement d'Avenir'' program, ANR-11-IDEX-0003-02, and by the Polish National Science Center (NCN) under the Grant No.: DEC-2011/01/B/ST8/07264 and  by the Polish National Center for Research and Development under the Grant No. PBS3/B9/34/2015.

\bibliography{main}

\begin{thebibliography}{10}

\bibitem{berkooz1993}
G.~Berkooz, P.~Holmes, and J.~L. Lumley.
\newblock The proper orthogonal decomposition in the analysis of turbulent
  flows.
\newblock {\em Annual review of fluid mechanics}, 25(1):539--575, 1993.

\bibitem{brunton2015}
S.~L. Brunton and B.~R. Noack.
\newblock Closed-loop turbulence control: Progress and challenges.
\newblock {\em Applied Mechanics Reviews}, 67(5):050801, 2015.

\bibitem{duriez2017}
T.~Duriez, S.~L. Brunton, and B.~R. Noack.
\newblock {\em Machine Learning Control-Taming Nonlinear Dynamics and
  Turbulence}.
\newblock Springer, 2017.

\bibitem{gautier2015}
N.~Gautier, J.-L. Aider, T.~Duriez, B.~Noack, M.~Segond, and M.~Abel.
\newblock Closed-loop separation control using machine learning.
\newblock {\em Journal of Fluid Mechanics}, 770:442--457, 2015.

\bibitem{hu2008a}
J.~Hu and Y.~Zhou.
\newblock Flow structure behind two staggered circular cylinders. part 1.
  downstream evolution and classification.
\newblock {\em Journal of Fluid Mechanics}, 607:51--80, 2008.

\bibitem{hu2008b}
J.~Hu and Y.~Zhou.
\newblock Flow structure behind two staggered circular cylinders. part 2. heat
  and momentum transport.
\newblock {\em Journal of fluid mechanics}, 607:81--107, 2008.

\bibitem{li2016}
R.~Li, D.~Barros, J.~Bor{\'e}e, O.~Cadot, B.~R. Noack, and L.~Cordier.
\newblock Feedback control of bimodal wake dynamics.
\newblock {\em Experiments in Fluids}, 57(10):158, 2016.

\bibitem{tutorial}
B.~Noack and M.~Morzy\'nski.
\newblock The fluidic pinball --- a toolkit for multiple-input multiple-output
  flow control (version 1.0).
\newblock Technical report, Institute of Combustion Engines and Transport,
  Poz\'nan University of Technology, 2017.

\bibitem{noack2003}
B.~R. Noack, K.~Afanasiev, M.~Morzy\'nski, G.~Tadmor, and F.~Thiele.
\newblock A hierarchy of low-dimensional models for the transient and
  post-transient cylinder wake.
\newblock {\em Journal of Fluid Mechanics}, 497:335--363, 2003.

\bibitem{parezanovic2016}
V.~Parezanovi{\'c}, L.~Cordier, A.~Spohn, T.~Duriez, B.~R. Noack, J.-P. Bonnet,
  M.~Segond, M.~Abel, and S.~L. Brunton.
\newblock Frequency selection by feedback control in a turbulent shear flow.
\newblock {\em Journal of Fluid Mechanics}, 797:247--283, 2016.

\end{thebibliography}
\bibliographystyle{abbrv}

\end{document}